\documentclass[a4paper,fleqn]{cas-sc}

\usepackage[authoryear,longnamesfirst]{natbib}
\usepackage{amsmath, hyperref}
\usepackage{graphicx,mathtools}
\usepackage[misc,geometry]{ifsym}
\usepackage{fontawesome}
\usepackage{lineno}
\usepackage{subcaption}
\usepackage{tablefootnote}
\usepackage{fontawesome}
\usepackage{supertabular}
\usepackage{multirow}
\usepackage{graphicx}
\usepackage{booktabs} 
\usepackage{amsfonts}
\usepackage{calligra}
\usepackage[xcolor]{changebar}
\usepackage{lipsum}

\def\tsc#1{\csdef{#1}{\textsc{\lowercase{#1}}\xspace}}
\tsc{WGM}
\tsc{QE}
\begin{document}
\let\WriteBookmarks\relax
\def\floatpagepagefraction{1}
\def\textpagefraction{.001}

\newcommand{\citetapos}[1]{\citeauthor{#1}'s \citeyearpar{#1}}

\shorttitle{Measuring Gender Bias in YouTube}    

\shortauthors{<short author list for running head>}  

\title [mode = title]{Measuring Gender Bias in Educational Videos: A Case Study on YouTube}  

\author[1,2]{Gizem Gezici}
\author[1]{Yucel Saygin}

\affiliation[1]{organization={Sabanci University},
            city={Istanbul},
%          citysep={}, % Uncomment if no comma needed between city and postcode
            country={Turkey}}

\affiliation[1]{organization={Huawei R\&D Center},
            city={Istanbul},
%          citysep={}, % Uncomment if no comma needed between city and postcode
            country={Turkey}}

\begin{abstract}
Students are increasingly using online materials to learn new subjects or to supplement their learning process in educational institutions. Issues regarding gender bias have been raised in the context of formal education and some measures have been proposed to mitigate them. However, online educational materials in terms of possible gender bias and stereotypes which may appear in different forms are yet to be investigated in the context of search bias in a widely-used search platform. As a first step towards measuring possible gender bias in online platforms, we have investigated YouTube educational videos in terms of the perceived gender of their narrators. We adopted bias measures for ranked search results to evaluate educational videos returned by YouTube  in response to queries related to STEM (Science, Technology,
Engineering, and Mathematics) and NON-STEM fields of education. Gender is a research area by itself in social sciences which is beyond the scope of this work. In this respect, for annotating the perceived gender of the narrator of an instructional video we used only a crude classification of gender into Male, and Female. Then, for analysing perceived gender bias we utilised bias measures that have been inspired by search platforms and further incorporated rank information into our analysis. Our preliminary results demonstrate that there is a significant bias towards the male gender on the returned YouTube educational videos, and the degree of bias varies when we compare STEM and NON-STEM queries. Finally, there is a strong evidence that rank information might affect the results.
\end{abstract}


%\item 
%\item 
%\item 

% Keywords

\maketitle

% Main text
%\section{}\label{}

% Use the style of numbering in square brackets.
% If nothing is used, default style will be taken.
%\begin{enumerate}[a)]
%\item 
%\item 
%\item 
%\end{enumerate}  

%\begin{itemize}
%\item 
%\item 
%\item 
%\end{itemize}  

%\begin{description}
%\item[]
%\item[] 
%\item[] 
%\end{description}  

\begin{comment}
% Figure
\begin{figure}[<options>]
	\centering
		\includegraphics[<options>]{}
	  \caption{}\label{fig1}
\end{figure}

\begin{table}[<options>]
\caption{}\label{tbl1}
\begin{tabular*}{\tblwidth}{@{}LL@{}}
\toprule
  &  \\ % Table header row
\midrule
 & \\
 & \\
 & \\
 & \\
\bottomrule
\end{tabular*}
\end{table}

\end{comment}

% Uncomment and use as the case may be
%\begin{theorem} 
%\end{theorem}

% Uncomment and use as the case may be
%\begin{lemma} 
%\end{lemma}


\section{Introduction}
\label{sec:intro}
Stereotypes are defined as beliefs regarding the characteristics, attributes, and behaviors of members of certain groups~\citep{Hilton1996Stereotypes}. Such beliefs, referring to society's stereotypes, often contain oversimplifications and prejudices about a specific group~\citep{Piatek-Jimenez2018InternationalFields}. Gender stereotypes begin to develop in early ages and these stereotypes about science, technology, engineering and mathematics (STEM) have severe consequences for motivation towards STEM fields~\citep{McGuire2020STEMCenters}. The early emerging gender stereotypes related to STEM are further strengthened in adolescence by the presence of male teachers and gender-imbalanced classrooms in STEM majors~\citep{Riegle-Crumb2017ShiftingGender}. A common stereotype is that STEM careers are for certain social groups such as European or American white males,~\citep{Bodzin2001BreakingStereotypes, Barman1997StudentsStudy} and this stereotype might signal to women and racial minority students that their group does not belong and is not successful in the STEM field~\citep{Good2012WhyMathematics}, thereby making them feel less welcoming, more insecure, and less motivated in STEM~\citep{London2011TheTransition}. Further, these stereotypes continue in the workplace and broader society, leading to the underrepresentation of woman in STEM fields~\citep{Piatek-Jimenez2018InternationalFields}. For instance, in the UK only 22\%, and in the US only 24\% of the STEM workforce is constituted by women~\citep{WISE, Noonan2017WomenUpdate}.

Gender stereotypes are a common source of bias that emerge when an individual or a group is systematically treated favourably or unfavourably, referring to~\emph{individual} or~\emph{group fairness} respectively and  there is a need to investigate gender representation in educational resources. In fact, the European Institute for Gender Equality states that gender stereotypes still exist in educational materials~\citep{EU}. There are some guidelines on how to evaluate diversity in educational materials, for example  Michigan in the United States issued a report in 2020 as a guidance for the experts in evaluating instructional materials in terms of bias~\citep{Center2020ToolsMaterials}.
These guidelines contain templates for scorecards to help the experts in their evaluation. Schools may try to implement the suggested guidelines and update/change their educational materials to mitigate the bias, but the problem may still persist since students are increasingly referring to online materials such as blogs, online educational websites, and YouTube videos. Nonetheless, gender seems to be an influential factor even in online search; it has been observed that university female students might be more readily to be affected by contexts than male students during online information seeking~\citep{zhou2014gender}. Thus, evaluating bias in online educational materials is very critical as well.
According to recent studies, YouTube has been declared as the world's second-most visited website and second-most used social platform worldwide~\citep{SalmanAslam2021YouTubeFacts}. In the third quarter of 2020 during the pandemic, YouTube had roughly 80\% market penetration in the UK, outperforming Facebook, WhatsApp, Instagram, and Twitter by number of active users and it had the highest reach among users aged 15 to 25 with 82\% of this demographic group~\citep{L.Ceci2021Social2021}. Although YouTube is popular as an entertainment medium, it has become a valuable alternative learning resource to written textual content such as blogs~\citep{Chintalapati2016ExaminingModel}.~\citet{Chtouki2012TheLearning} report  a study showing that visual explanations help students to understand and remember the complex concepts much better. These studies justify that YouTube has been a widely-used platform as well as an effective tool for improving student's learning and engagement.

YouTube states that they audit their machine learning systems to avoid cases leading to gender discrimination~\citep{YouTube}. However, this does not guarantee that the returned videos are not biased towards a specific gender. In this study, our aim is to  investigate educational videos returned by YouTube in terms of possible gender bias via objective measures. We focus on~\emph{group fairness}  since we investigate if the online  materials are affected by societal stereotypes about gender in the context of education. Moreover in~\emph{group fairness}, we specifically focus on~\emph{statistical parity}, \emph{demographic parity} or more generally known as~\emph{equality of outcome}, i.e. given a population divided into groups, the groups in the output of the system should be equally represented. In the scope of this work,~\emph{equality of outcome} is a more appropriate standard since we require~\emph{equal} gender representations in results.
Our main aim in this study is to detect bias with respect to~\emph{equality of outcome} using the~\emph{perceived} gender of narrators in videos returned by YouTube in response to educational queries comprising of keywords regarding some educational field. For this purpose, we use educational queries that are derived from the course modules %for the corresponding majors areas of study in 
of STEM and NON-STEM fields.

Our contributions  can be summarised as follows:

\begin{enumerate}

\item We present~\emph{two new measures of bias} which are explained in Section~\ref{sec:measuresbias} in detail that treat our two protected groups equal, and generate bias values which are symmetric and easy to interpret.

\item We implement our bias measures to \emph{investigate possible perceived gender bias} for educational searches in YouTube about different majors from STEM and NON-STEM fields.

\item We also \emph{compare the relative bias} for educational queries in YouTube from various majors from STEM and NON-STEM fields.

\item We further incorporate rank information into the bias analysis to investigate if various rank values affect first the~\emph{existence of bias}, then~\emph{difference in magnitude of bias} between STEM and NON-STEM fields as well as in the same field.

\end{enumerate}

\section{Related Work}
\label{sec:related_work}
Before presenting our methodology for evaluating~\emph{perceived} gender bias in YouTube video results, we review the prior work related first to fair ranking evaluation, and second to search bias quantification.

\subsection{Fairness in Ranking}
\label{sec:faireval_related}
Many recent studies have investigated two different notions of fairness as~\emph{individual fairness} and~\emph{group fairness} in ranked outputs. Individual fairness requires that similar individuals should be treated similarly, whereas group fairness requires that the disadvantaged group be treated similar to the advantaged group or the entire population~\citep{Dwork2012FairnessAwareness}. Formal definitions of group fairness is composed of~\emph{statistical parity}, \emph{demographic parity} or more generally known as~\emph{equality of outcome}~\citep{Dwork2012FairnessAwareness}, and equality of opportunity~\citep{Hardt2016equality}. Various fairness evaluation measures in ranked results have been proposed in the literature.~\citet{Yang2016MeasuringOutputs} propose a synthetic ranking generation procedure and three measures inspired by utility-based information retrieval (IR) evaluation measures that are related to normalized Discounted Cumulative Gain (nDCG) in the scope of statistical parity.~\citet{Zehlike2017FAIR:Algorithm} present an algorithm based on~\citet{Yang2016MeasuringOutputs}'s work using statistical tests to produce and evaluate fair rankings. These measures are difficult to use in practice since they rely on a normalisation term which is computed stochastically, i.e. the notion of an ideal list for normalisation similar to nDCG, and they do not consider relevance which is crucial in search settings.~\citet{Gezici2021EvaluationResults} then improve these measures by addressing their limitations in the context of search results. In this work, we focus on~\emph{perceived} gender bias and propose two new bias measures that address the shortcomings of the measures proposed by~\citet{Gezici2021EvaluationResults}. Our measures compute percentage scores of representation and exposure bias which are easy to interpret to assess~\emph{equality of outcome} in search results.
Unlike~\cite{Gezici2021EvaluationResults}, in  both of our bias measures, we take into account the YouTube video results only annotated with~\emph{male} and~\emph{female} gender labels meaning that we compute the score of a male/female gender over the sum of male and female scores using the two proposed bias measures. In this way, difference in computed metric values which shows the inequality between the~\emph{perceived} gender groups becomes more significant and the scores of each gender group are symmetric around $0.5$, which is the desired case. Additionally, in our exposure measure with logarithmic weighting we require a normalisation term for interpretable results, yet we do not include the notion of an ideal list that was used by the researchers in~\cite{Yang2016MeasuringOutputs} since its definition differs with each ranked list in our dataset which is not practical to compute.

\citet{Geyik2019Fairness-AwareSearch} present two measures in the scope of~\emph{equality of opportunity}, one of these measures is based on~\citet{Yang2016MeasuringOutputs}'s measures and the other one compares the representation of a gender group in a given list with respect to the entire population.~\citet{Lipani2021TowardsBias} propose a measure which compares the representation of a categorical sensitive attribute in result documents with all the documents indexed by the search system.~\citet{Gomez2021TheSystems} propose two measures which evaluate a given ranked list in terms of representation, i.e. proportion without including rank information, and exposure, i.e. incorporating stronger rank information with logarithmic weighting; both measures compare the given list with the full dataset. Thus, unlike us,~\citet{Gomez2021TheSystems, Geyik2019Fairness-AwareSearch, Lipani2021TowardsBias} propose measures to achieve~\emph{equality of opportunity}, yet the measures are computed on search results in~\cite{Geyik2019Fairness-AwareSearch, Lipani2021TowardsBias}, while in~\cite{Gomez2021TheSystems} they are applied in recommendation settings. As discussed in Section~\ref{sec:intro}, our methodology implements~\emph{equality of outcome} in search settings instead, since we analyse gender equality in the context of online educational materials. This requires that narrators should be treated~\emph{equally} regardless of their gender which is a sensitive attribute that should not affect the decision of the ranking algorithm.

\subsection{Bias Quantification in Search}
\label{sec:biassearch_related}
Although the algorithms of search platforms are not transparent to researchers, auditing techniques can help us to reveal biases to which users are being exposed without even being aware of the existence of bias. Moreover, it was shown that people are more susceptible to bias when they are unaware of it~\citep{Bargh2001TheGoals}. Thus, there have been several attempts to audit bias on web search for a variety of topics, for instance~\citet{Diakopoulos2018ICandidate, Robertson2017SuppressingSEME, Epstein2015TheElections, Robertson2018AuditingSearch, Robertson2018AuditingPages, Hu2019AuditingSnippets} particularly focus on~\emph{partisan bias} in search results. Apart from partisan bias, recent studies have investigated~\emph{gender bias} in search.~\citet{Chen2006PositionAssessment} examine gender bias in various resume search engines using a regression analysis in the context of individual and group fairness and found that there is a significant and consistent group unfairness against female candidates. Using a similar approach,~\citet{Hannak2017bias} investigate~\emph{perceived gender} and~\emph{race} bias in two prominent online freelance websites to capture correlations between the profile features of workers and their reviews/ratings as well as their search rank position. They found that~\emph{perceived gender} and~\emph{race} bias are negatively correlated with search rank in one of these freelance websites.~\citet{Kay2015UnequalOccupations} investigate the gender bias in image search results for a variety of occupations by only applying statistical significance tests without modelling the bias problem in the context of search and revealed that image search results exaggerate gender stereotypes and display the minority gender rather unprofessionally. Similarly,~\citet{singh2020female} examine the image-based representation of highly gender-discriminated professions, e.g. nurse, computer programmer, in four digital platforms by simply comparing the ratio of male and female images in those platforms with the national labor statistics and found that women are largely underrepresented. Likewise,~\citet{Otterbacher2017CompetentResults} inspect gender stereotypes by directly computing the gender proportions in image search results returned by Bing without proposing any specific measures and showed that photos of women are more often retrieved for `emotional' and similar traits, whereas men for `rational' and related traits. In a follow-up work, by using a regression analysis the authors showed that sexist people are less likely to detect and report gender biases in search results~\citep{Otterbacher2018InvestigatingSexism}. In addition to these studies which view gender bias in the context of search, some researchers also examine the relationship between how the course is displayed, i.e. the course is presented with a gender-inclusive photo, descriptions that contain more negative sentiment etc. which are the psychological cues, and the enrollment/engagement of different genders to STEM courses in online learning platforms~\citep{kizilcec2019psychologically, kizilcec2020identifying, brooks2018gender}.

Users typically pay more attention to top positions in a ranked list of search results which is called position bias and this phenomenon leads users to click those top positions with greater probability~\citep{Joachims2005AccuratelyFeedback}. Therefore, if search results are biased then users will be affected due to search engine manipulation effect (SEME)~\citep{Epstein2015TheElections} and the impact is high if top positions are more biased. Since users tend to click the top positions with higher probability, this implicit user feedback will be logged, then fed to the ranking algorithm which will probably cause users to be exposed to an even higher bias -- societal biases will be reinforced in the search results. Thus, even if the bias comes from the data itself, search platforms should still be responsible for mitigating it. As stated by~\citet{culpepper2018research}, an information retrieval system should be fair, accountable, and transparent. For preventing the severe consequences of bias in society, the first step is to reveal it, which is our main focus in this work, thereby further alerting users which could be effective in suppressing search engine manipulation effect~\citep{Robertson2017SuppressingSEME}.

\section{Preliminaries and Research Questions}
\label{sec:prelim}

We assume a scenario where a query such as "Gravity" or "Python Programming" is issued to YouTube and a result page with a list of videos is returned in response to the query. We call such queries educational queries and we use the abbreviation of~\emph{YVRP} for the YouTube video result page for a given query throughout the paper. We consider the gender of the narrator in this work and our first task is to label the videos with respect to the ~\emph{perceived}  gender of the narrator. A~\emph{perceived} gender label can have the following values:

\emph{male}, % ($+$), 
\emph{neutral}, % ($\cdot$), 
\emph{female}, % (-) or 
\emph{not-relevant}, and % ($\times$).
\emph{N/A} with respect to the viewer's overall perception and their meanings are as follows:
\begin{itemize}
\vspace{-0.5em}
\item\textbf{male} ($G_m$) If  the video is mostly narrated by people whose gender is perceived as male;
\item\textbf{neutral} ($G_{neut}$) When the video does not favour either male or female gender in narration. Therefore, the video does not help the viewer to infer any gender dominance;
\item\textbf{female} ($G_f$) If the video is mostly narrated by people whose gender is perceived as female; 
\item\textbf{not-relevant} ($G_{not\_rel}$) when the video is not-relevant with respect to the educational query;
\item\textbf{N/A} ($G_{N/A}$) When the annotation is not applicable for the video -- the video is not in English or it has been removed from the system, or there is no narrator.
\end{itemize}

A \emph{YVRP} contains 12 video links. In Figure~\ref{fig:fig_rankedlist} we provide different ranked lists of labelled results. In Figure~\ref{fig:fig_rankedlist}~(a) the perceived gender of all the narrators is labelled as male which demonstrates a clear bias. In Figure~\ref{fig:fig_rankedlist}~(b) and (c) half of the perceived genders are male and half of them are female however in Figure~\ref{fig:fig_rankedlist}~(b) the top 6 ranked videos are labelled as male while in Figure~\ref{fig:fig_rankedlist}~(c) the top 6 are labelled as female. In Figure~\ref{fig:fig_rankedlist}~(d) there is no obvious bias, while in the videos of Figure~\ref{fig:fig_rankedlist}~(e) we have neutral, not-relevant, and N/A labels which further complicate the bias evaluation.  
We need to issue many and different queries and evaluate the results for possible bias before we reach a conclusion. The first research question we aim to answer is:

\begin{description}
    \item[RQ1:] On a~\emph{perceived} male-female binary gender space, does YouTube return~\emph{biased} \emph{YVRP}s in response to various educational queries? % towards educational queries?related to STEM and NON-STEM fields? 
\end{description}

There are different fields of education which are broadly categorized as STEM and NON-STEM where the number of female students in some STEM fields has been considerably less than the males. Our second  research question is: 

\begin{description}
 \item[RQ2:] Is there a \emph{significant} difference in~\emph{perceived} gender bias in \emph{YVRP}s returned in response to STEM vs. NON-STEM educational queries? % Do STEM and NON-STEM fields show \emph{significantly different} magnitude of~\emph{perceived} gender bias from each other?% towards educational queries in YouTube?   
\end{description}

We provide bias evaluation measures that take into account the rank of the results. One of the measure looks at the top $n$ results in comparison to the rest of the 12 results, where $n$ is the cut-off value. For example in the videos of Figure~\ref{fig:fig_rankedlist}~(d) when we have a cut-off value of 3, there will be a significant difference in bias in top 3 vs top 12. Our third research question is:

\begin{description}
    \item[RQ3:] Do different cut-off values affect the existence of~\emph{perceived} bias and magnitude of bias difference between STEM and NON-STEM fields? 
    %%%%%Rephrase the esearch question 3!!!!
\end{description}

Finally, in addition to the impact of different cut-off values on the existence of~\emph{perceived} bias in STEM and NON-STEM fields, we further examine how the cut-off values influence on the magnitude of bias in each field. Our last research question is:

\begin{description}
    \item[RQ4:] Do different cut-off values affect the magnitude of~\emph{perceived} bias of STEM and NON-STEM fields?
\end{description}

\section{Gender Bias Evaluation Methodology}
\label{sec:gender_framework}
In this section we describe our~\emph{perceived} gender bias evaluation methodology with the gender binary assumption. We present two  measures of bias  and a protocol to identify possible bias with respect to those measures.

\subsection{Measures of Bias}
\label{sec:measuresbias}

Let $Q$ be the set of educational queries about major areas of study in STEM and NON-STEM fields. 
When a query $q \in Q$ is issued to YouTube, YouTube returns a \emph{YVRP} $r$. We define the~\emph{perceived} gender of the $i$-th retrieved video $r_i$ with respect to $q$ as $j(r_i)$. For reference, Table~\ref{table:notation} shows a summary of all the symbols, functions and labels used throughout the paper.

For satisfying the group fairness criterion of~\emph{equality of outcome}, male and female genders should be~\emph{equally} represented in the retrieved YouTube videos.
In the scope of~\emph{perceived} gender bias analysis, we can mention the existence of bias in a ranked list of videos retrieved by YouTube, if the~\emph{perceived} gender representation \emph{significantly deviates} from~\emph{equal} representation. Thus, we need to measure the difference between the representation of two genders, which we indicate here as male and female.

\begin{table*}
\centering
\caption{Symbols, functions, and labels used throughout the paper}
\setlength{\tabcolsep}{0.8pt}
\renewcommand{\arraystretch}{0.6}
\begin{tabular}{ l l }
\hline
\hline
{Symbols}\\
\hline
\hline
%$\sS$   & set of fields. \\
%$s$     & a field $s \in \sS$ (STEM or NON-STEM). \\
$Q$   & set of queries. \\
$q$     & a query $q \in Q$. \\
%$d$     & a document. \\
$r$     & a ranked list of the given \emph{YVRP} (list of retrieved videos). \\%(list of documents).
$r_i$   & the video in $r$ retrieved at rank $i$.\\
$|r|$ & size of $r$ (number of videos in the ranked list). % = [r_1, r_2, \dots, r_{|r|}]$.
\\
$n$     & number of videos considered in $r$ %by $f$ 
(cut-off).\\
$n_{mf}$     & number of videos in $r$ which are annotated as male or female (excluding neutral, not-relevant, and N/A).\\ %by $f$\\
%$G$     & a set of groups $g_i$ ($g_1 \cup g_2 \cup \dots \cup g_n = G$). \\
%$g_i$   & a group $g_i \in G$. \\ \hline
%\hline
\hline
\hline
{Functions}\\
\hline
\hline
%$s(q)$                  & returns $r$ given a query ($q$). \\ 
$j(r_i)$                & returns the label associated to $r_i$.\\
%$j(q)$                  & returns the label associated to $q$.\\
$f(r)$                  & an evaluation measure for \emph{YVRP}s.\\
%$f_g(r)$                & like $f$ but for $g$.\\
%$f_{-}(r)$             & like $f$ but for the against label.\\
%$[\text{Condition}]$    & returns 1 if the condition is true and 0 otherwise.\\
\hline
\hline
{Labels}\\
\hline
\hline
%$g_1$  & protected group. \\
%$g_2$  & unprotected group. \\ \hline
$G_m$ & \emph{perceived} male gender. \\
$G_{neut}$ & \emph{perceived} neutral gender. \\
$G_f$ & \emph{perceived} female gender. \\ 
$G_{not\_rel}$ & not-relevant wrt a query. \\
$G_{N/A}$ & N/A - gender annotation is not applicable. \\
%$+$ & label  .\\
%$-$ & against label.\\
\hline
\hline
\end{tabular}
%\vspace{-1.3em}
\label{table:notation}
\vspace{-1.4em}
\end{table*}

Formally, we measure the \emph{perceived gender} bias in a \emph{YVRP} $r$ as follows:
\begin{linenomath*}
\begin{equation}\label{eq:delta_bias}
    \Delta_f(r) = f_{G_{m}}(r) - f_{G_{f}}(r)
\end{equation}
\end{linenomath*}
where $f$ is a function that measures the likelihood of $r$ in satisfying the information need of the user about the~\emph{perceived} gender of male ($G_{m}$) and female ($G_{f}$). When $\Delta_f(r)$ = 0 we consider that $r$ to be bias-free. When $\Delta_f(r)$ > 0, the YVRP is biased towards male ($G_{m}$), with maximal bias when $\Delta_f(r)$ = 1. When $\Delta_f(r)$ < 0, then the YVRP is biased towards female ($G_{f}$), with maximal bias when $\Delta_f(r)$ = -1.

For the function $f(r)$, we propose two~\emph{novel} bias measures in the scope of~\emph{equality of outcome}. Please note that only the videos annotated with the~\emph{perceived} gender labels of male ($G_{m}$) and female ($G_{f}$), that are relevant to the query, are taken into account. We discard the videos for which $j(r_i)$ returns neutral ($G_{neut}$), not-relevant ($G_{not-rel}$), or N/A ($G_{N/A}$). Note that, $j(r_i)$ returns the label of video $r_i$ specifying its gender group. Based on this, $[j(r_i) = G_{m}]$ refers to a conditional statement which returns 1 if the video $r_i$ is annotated as the member of $G_{m}$ and 0 otherwise. The  two new measures of~\emph{representation} and~\emph{exposure} are denoted by~$\mathcal{R}ep@n$ and $\mathcal{E}xp@n$ respectively. The measure of $\mathcal{R}ep@n$ deals with the bias in~\emph{gender proportion}, while $\mathcal{E}xp@n$ aims to reveal the bias caused by~\emph{exposure effects}, i.e. attention received by ranked items. Our first measure of bias, $\mathcal{R}ep@n$ which is interpreted with respect to the~\emph{perceived} gender label of~\emph{male} as follows:

\begin{linenomath*}
\begin{equation}\label{eq:rep}
    \mathcal{R}ep_{G_{m}}@n = 
    \frac{1}{n_{mf}}\sum_{i=1}^n
    [j(r_i) = G_{m}]
\end{equation}
\end{linenomath*}

Note that $\mathcal{R}ep_{G_{f}}@n$ is computed in the same way. Substituting~\eqref{eq:rep} in~\eqref{eq:delta_bias} we obtain:

\begin{linenomath*}
\begin{equation}\label{eq:bias_rep}
    \Delta_{\mathcal{R}ep@n}(r) = 
    \frac{1}{n_{mf}}{\sum_{i=1}^n\left([j(r_i) = G_{m}] - [j(r_i) = G_{f}]\right)}
    %\frac{1}{n_{mf}}\sum_{i=1}^n 
    %\left([j(r_i) = G_{m}] - [j(r_i) = G_{f}]]\right)
\end{equation}
\end{linenomath*}

Although the first bias measure of $\mathcal{R}ep@n$ is very intuitive, it is insensitive to the rank positions since all the search results in the first $n$ documents contribute to the bias score equally, regardless of their rank positions. Thus, we  propose the second measure of $\mathcal{E}xp@n$ to address this issue by defining a discount function based on rank which helps us to include a strong concept of ranking information in our bias analysis. Our logarithmic discounting method is inspired by the weighted discount mechanism of nDCG which is a widely used utility-based IR metric. 
Our new measure of $\mathcal{E}xp@n$ computes the exposure percentage which belongs to each gender group of male and female. The proposed measure of $\mathcal{E}xp@n$ which is interpreted with respect to the~\emph{perceived} gender label of~\emph{male} as follows:

\begin{figure}[!t]
\caption{Example ranked lists of YouTube results for a query.}
\centering
%\hspace{-1.8em}
    \begin{subfigure}{0.15\textwidth}
        \centering
        \includegraphics[width=0.5\linewidth]{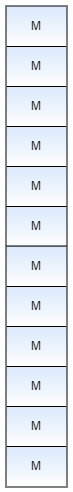}
        \subcaption{}
        \label{fig:fig1}
\end{subfigure}
%\hspace{1.8em}
\begin{subfigure}{0.15\textwidth}
        \centering
        %\hspace{-1.8em}
        \includegraphics[width=0.5\linewidth]{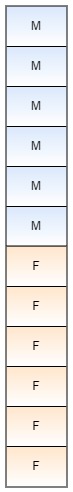}
        \subcaption{}
    \label{fig:fig2}
\end{subfigure}
\begin{subfigure}{0.15\textwidth}
        \centering
        %\hspace{-1.8em}
        \includegraphics[width=0.5\linewidth]{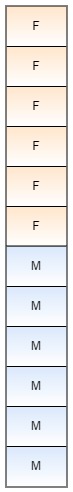}
        \subcaption{}
    \label{fig:fig3}
\end{subfigure}
\begin{subfigure}{0.15\textwidth}
        \centering
        %\hspace{-1.8em}
        \includegraphics[width=0.5\linewidth]{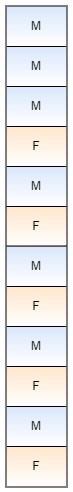}
        \subcaption{}
    \label{fig:fig4}
\end{subfigure}
\begin{subfigure}{0.15\textwidth}
        \centering
        %\hspace{-1.8em}
        \includegraphics[width=0.5\linewidth]{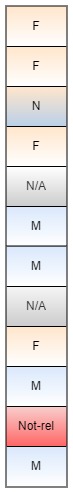}
        \subcaption{}
    \label{fig:fig5}
\end{subfigure}
\label{fig:fig_rankedlist}
%\vspace{-1.2em}
\end{figure}

\begin{linenomath*}
\begin{equation}\label{eq:dcg_exposure}
    \mathcal{E}xp_{G_{m}}@n = 
    \sum_{i=1}^{n}\frac{1}{\log(i + 1)}(\frac{[j(r_i) = G_{m}]}{[j(r_i) = G_{m}] + [j(r_i) = G_{f}]})
\end{equation}
\end{linenomath*}

Note that $\mathcal{E}xp_{G_{f}}@n$ is computed in the same way. Substituting Eq.~\eqref{eq:dcg_exposure} to Eq.~\eqref{eq:delta_bias} we obtain:

\begin{linenomath*}
\begin{equation}\label{eq:bias_exposure}
    \Delta_{\mathcal{E}xp@n}(r) = 
    \sum_{i=1}^{n}\frac{1}{\log(i + 1)}(\frac{[j(r_i) = G_{m}] - [j(r_i) = G_{f}]}{[j(r_i) = G_{m}] + [j(r_i) = G_{f}]})
\end{equation}
\end{linenomath*}
The scores of the proposed measures are easy to interpret, for a given ranked list the scores of each gender group sum up to 1. If we interpret the bias scores with respect to the~\emph{equal} representation using $\mathcal{R}ep@n$, then we can infer which gender group is more/less represented than the desired representation. Same applies to the exposure measure, $\mathcal{E}xp@n$.
%based on which we can conclude that a gender group has higher/lower exposure with respect to the  desired case of ~\emph{equal} exposure. 
For interpreting the results, if we subtract the value of $0.5$ which is the desired case, from the measure scores, then the remaining bias scores of male and female are symmetric. Same holds for the exposure measure. 
These bias measures are calculated for the sample ranked lists in Figure~\ref{fig:fig_rankedlist} as follows:

\begin{itemize}
\item In Figure~\ref{fig:fig_rankedlist} (a), the perceived gender of all the narrators are labelled as male. For this ranked list $r$, $\mathcal{R}ep_{G_{m}}@12$ = $1$, whereas $\mathcal{R}ep_{G_{f}}@12$ = $0$, thereby $\Delta_{\mathcal{R}ep@12}(r)$ = $1$ which is the maximal bias. Same exposure bias score is computed for this ranked list.

\item In Figure~\ref{fig:fig_rankedlist} (b), half of the perceived genders are male and the top 6 ranked videos are labelled as male. For this ranked list $r$,
$\mathcal{R}ep_{G_{m}}@12$ = $0.5$ and $\mathcal{R}ep_{G_{f}}@12$ = $0.5$, thus $\Delta_{\mathcal{R}ep@12}(r)$ = $0$ indicating no representation bias. On the other hand, $\mathcal{E}xp_{G_{m}}@12$ = $0.65$ and $\mathcal{E}xp_{G_{f}}@12$ = $0.35$, thus $\Delta_{\mathcal{E}xp@12}(r)$ = $0.30$. Since the first measure only looks at the proportion of gender groups in the given ranked list without taking into account the rank information, no representation bias is observed. However, using the second measure which uses rank information with a logarithmic discount function, it can be seen that there exists an~\emph{exposure bias} towards the~\emph{male} gender since $\Delta_{\mathcal{E}xp@12}(r)$ > 0.

\item In Figure~\ref{fig:fig_rankedlist} (c), again half of the perceived genders are male however unlike Figure~\ref{fig:fig_rankedlist} (b), the top-6 ranked videos are labelled as female. For this ranked list $r$, our representation bias measure computes the same scores with the ranked list $r$ in (b) as $\mathcal{R}ep_{G_{m}}@12$ = $0.5$ and $\mathcal{R}ep_{G_{f}}@12$ = $0.5$, thus $\Delta_{\mathcal{R}ep@12}(r)$ = $0$ showing no representation bias. On the other hand, $\mathcal{E}xp_{G_{m}}@12$ = $0.35$, whereas $\mathcal{E}xp_{G_{f}}@12$ = $0.65$, thus $\Delta_{\mathcal{E}xp@12}(r)$ = $-0.30$. As with the ranked list in (b), there does not exist representation bias since $\Delta_{\mathcal{R}ep@12}(r)$ = $0$, while there exists an~\emph{exposure bias} towards the~\emph{female} gender since $\Delta_{\mathcal{E}xp@12}(r)$ < 0.

\item In Figure~\ref{fig:fig_rankedlist} (d), the ranked list $r$ contains almost the same number of male and female perceived gender labels, yet in the top-3 all the narrators are labelled as male. For this ranked list $r$, $\mathcal{R}ep_{G_{m}}@12$ = $0.58$, whereas $\mathcal{R}ep_{G_{f}}@12$ = $0.42$, thereby $\Delta_{\mathcal{R}ep@12}(r)$ = $0.16$ indicating a low representation bias, very close to the bias-free case. On the other hand, $\mathcal{E}xp_{G_{m}}@12$ = $0.67$, whereas $\mathcal{E}xp_{G_{f}}@12$ = $0.33$, thus $\Delta_{\mathcal{E}xp@12}(r)$ = $0.34$ revealing a higher bias in exposure than the representation. In this case, a higher exposure bias is observed bias since the exposure measure takes into account the rank information. Moreover, if bias is measured for different cut-off values, then various level of bias might be obtained in representation and exposure. For instance, let's only look at the top-3 positions in the ranked list $r$, where $n$ = $3$. Then, $\mathcal{R}ep_{G_{m}}@3$ = $1$ and $\mathcal{R}ep_{G_{f}}@3$ = $0$, thus $\Delta_{\mathcal{R}ep@3}(r)$ = $1$ which is the maximal bias in comparison to the full list. Similarly, $\Delta_{\mathcal{E}xp@3}(r)$ = $1.0$ which is the maximal bias for exposure as well.

\item In Figure~\ref{fig:fig_rankedlist} (e), the ranked list $r$ contains the same number of male and female perceived gender labels, yet female gender appears more in the top positions. Additionally, unlike the previous lists this ranked list contains neutral, not-relevant, and N/A. Since the perceived gender labels of neutral, not-relevant, and N/A do not contribute to detect gender bias, these labels are not included in our computations. Therefore, for this ranked list $r$, $\mathcal{R}ep_{G_{m}}@12$ = $0.5$ and $\mathcal{R}ep_{G_{f}}@12$ = $0.5$, thus $\Delta_{\mathcal{R}ep@12}(r)$ = $0$, no representation bias. This is because if the aforementioned labels are discarded, the ranked list turns into the same lists as in (b) and (c). On the other hand, $\mathcal{E}xp_{G_{m}}@12$ = $0.35$, while $\mathcal{E}xp_{G_{f}}@12$ = $0.65$, thus $\Delta_{\mathcal{E}xp@12}(r)$ = $-0.30$ which indicates that there is an exposure bias. Similar to the ranked lists in (b) and (c), no representation bias is observed since $\Delta_{\mathcal{R}ep@12}(r)$ = $0$, whereas there exists an~\emph{exposure bias} towards the~\emph{female} gender since $\Delta_{\mathcal{E}xp@12}(r)$ < 0.
\end{itemize}

These computations demonstrate that both of the proposed measures are necessary since they provide different types of information for the analysis. Moreover, the findings show that the magnitude of bias we observe differs with different cut-off values ($n$), therefore we will use various values of $n$ in Section~\ref{sec:experimental}. Since users tend to pay more attention to the top positions in search results, the impact of higher bias in these positions could be more severe in the scope of gender equality. In addition to the bias computations, for interpreting the bias scores in representation, we can subtract the value of $0.5$ to obtain the relative representations of male and female in a ranked list $r$. For instance, in Figure~\ref{fig:fig_rankedlist} (d), we obtained $\mathcal{R}ep_{G_{m}}@12$ = $0.58$ and $\mathcal{R}ep_{G_{f}}@12$ = $0.42$ for male and female gender labels respectively. Thus, if we subtract the value of $0.5$, we obtain $0.08$ for male and $-0.08$ for female which means that the male gender is represented 8\% more, and the female gender is represented 8\% less than the~\emph{equal} representation. Similarly, we obtained $\mathcal{E}xp_{G_{m}}@12$ = $0.67$ and $\mathcal{E}xp_{G_{f}}@12$ = $0.33$ for male and female gender labels respectively and if we subtract the value of $0.5$, we obtain $0.17$ for the male and $-0.17$ for female genders. From this, we can infer that the male gender obtains 17\% more exposure while the female gender gets 17\% less exposure than the desired case.

After the computation of representation and exposure bias scores, we compute the mean bias (MB) and mean absolute bias (MAB) of these measures over a set of queries in the dataset to aggregate the bias results. MB score of STEM field computes a mean value over all the STEM queries' scores for the corresponding measure, whereas MAB computes a mean value over all the absolute value of the measure scores for the STEM queries. Note that MB shows towards which~\emph{perceived} gender the results are biased and MAB solves the limitation of MB if different queries have bias contributions with opposite signs and cancel each other out. Thus, MB and MAB measures are complementary for aggregating the results and interpreting those results in a proper way.

Please note that in the scope of this work, we assign the gender label of a given video merely based on the narrators'~\emph{perceived} gender and we make gender binary assumption. However, our definitions and thereby our measures of bias can easily be applied to studies where the gender label is defined in a more refined manner. Labels can also be assigned based on the male/female dominance, similar to the viewpoints presented by~\citet{Draws2021ThisTopics}. Yet, in the scope this work, we accept the~\emph{perceived} gender label as binary for the preliminary results. Moreover, the proposed measures are also suitable for studies that use similar categorical features like age, education, ethnicity, or geographic location~\citep{Lipani2021TowardsBias} and seek for~\emph{demographic parity} specifically, in search settings.

\subsection{Quantifying Bias}
\label{sec:protocol}
Using the measures of bias defined in Section~\ref{sec:measuresbias}, we quantify the~\emph{perceived} gender bias of STEM and NON-STEM fields in~\emph{YVRP}s returned in response to the educational queries in various majors, and we compare them. We describe our methodology for quantifying~\emph{perceived gender} bias.

\begin{itemize}
\item \textbf{Collecting~\emph{YVRP}s.}
We obtained the educational queries
issued for searching in YouTube from~\emph{TheUniGuide}~\footnote{https://www.theuniguide.co.uk/}. TheUniGuide is a free university advice service which is part of The Student Room~\footnote{Free student discussion forum in UK} that helps students make more informed decisions about their higher education choices.
We submitted each query to
YouTube using a UK proxy in~\emph{incognito mode} and crawled the top-12 video results returned by YouTube. Note that the data collection process was done in a controlled environment such that the queries are sent to YouTube by avoiding long time-lags. After the crawling of all the~\emph{YVRP}s related to the majors in both STEM and NON-STEM fields, we labelled them. We annotated the~\emph{perceived} gender label of each video with respect to the educational queries by analysing the gender(s) of the narrator(s) from the viewer's perspective. 
\item \textbf{Bias Evaluation.}
We compute the bias scores for every~\emph{YVRP} with two new bias measures with three different cut-off values: $\mathcal{R}ep@n$ and $\mathcal{E}xp@n$ for $n$ = $3$, $6$, $12$. We then aggregate the results using the MB and MAB. Then, we first examine the existence of bias for each field, further compare the bias results of STEM and NON-STEM fields with different measures and cut-off values. Finally, we investigate the impact of different cut-off values on bias scores of STEM and NON-STEM fields.

\item \textbf{Statistical Analysis.}
To identify whether the bias measured is not due to noise,
we compute a one-sample t-test: 
the null hypothesis is that no difference exists and that the true mean is equal to zero. Note that since our sample size is sufficiently large (> 30), according to the central limit theorem the sampling distribution is considered normal~\citep{Kwak2017CentralStatistics}.
If this hypothesis is rejected, hence there is a significant difference and we claim that the~\emph{YVRP}s of the evaluated field, STEM or NON-STEM is biased. We compare the difference in bias measured across the two fields using a two-tailed independent t-test since we have a different set of queries for each field: the null hypothesis is that the difference between the true means of the two independent groups is equal to zero. If this hypothesis is rejected, hence there is a significant difference, we claim that there is a difference in bias between the two fields. The acceptance or rejection of the null hypothesis is fulfilled based on the p-values. Note that before applying the two-tailed independent t-test, we do not check if the two samples have equal or unequal variances, but rather directly use the independent t-test with unequal variances~\citep{delacre2017psychologists}. Nonetheless, in the context of our analysis, it seems that independent t-test with equal or unequal variances do not make a noticeable difference in p-values based on our initial analysis.
In addition to the statistical significance, namely p-values, we report effect sizes using Cohen's d. Statistical significance in our analysis helps us examine whether the findings show systematic bias or they are the result of noise, whereas effect sizes provide information about the magnitude of the differences which makes both p-values and effect sizes complementary for the interpretation of our results.

Apart from these, to investigate the effect of different cut-off values on bias results in the same field, STEM or NON-STEM, we compute a two-tailed paired t-test since in this analysis we examine the same query set only with different cut-off values. Moreover, we further apply Bonferroni correction~\citep{sedgwick2012multiple} for multiple hypothesis testing since there are 24 hypotheses in total in the context of cut-off value analysis. Thus, without the Bonferroni correction, with the significance level, $ \alpha = .05$ and 24 hypotheses, the probability of identifying at least one significant result due to chance is around $0.71$ which means that the results could be misleading. Based on these, we further apply the Bonferroni correction for more reliable results in the scope of the cut-off value analysis in Section~\ref{sec:results}. Note that for the significance level where $\alpha = .05$, and with the Bonferroni correction new $\alpha = .02$. Thus, Bonferroni correction rejects the null hypothesis for each p-value ($p_i$) if $p_i$ <= $.02$ instead of $.05$.

\end{itemize}

\begin{table}[!t]
%\vspace{1em}
\setlength{\tabcolsep}{2.3pt}
\renewcommand{\arraystretch}{2.6}
\centering
\caption{All the course modules of TheUniGuide we used as the user queries for the main study.}
\resizebox{\columnwidth}{!}{%
\begin{tabular}{cc|c|ccc|c|c}
    \hline
    \textbf{STEM} & \multicolumn{3}{c}{~\textbf{Course Modules}} & \textbf{NON-STEM} & \multicolumn{3}{c}{~\textbf{Course Modules}}\\
    \hline
    \multirow{2}{*}[-4em]{\textbf{Biology}} & Biochemistry & \shortstack{Evolution and \\ biodiversity} & \shortstack{Marine and \\ terrestrial ecology} & \multirow{2}{*}[-4em]{\textbf{\shortstack{English Language \\ and Literature}}} & \shortstack{Explorations \\in literature} & \shortstack{Chaucer: texts, \\ contexts, conflicts} & \shortstack{Shakespeare in performance \\~\textbf{~\textcolor{red}{english language and literature}}}\\
    \cline{2-4}
    \cline{6-8}
    & \shortstack{Plant  science \\~\textbf{~\textcolor{red}{in biology}}} & \shortstack{Human \\ physiology} & \shortstack{Habitat ecology \\~\textbf{~\textcolor{red}{in biology}}} & & \shortstack{Renaissance \\literature} &  \shortstack{Modernist \\fiction} & \shortstack{Creative \\writing: drama} \\
    \cline{2-4}
    \cline{6-8}
    & \shortstack{Environmental \\ issues} & \shortstack{Molecular methodology \\ for biologists} & \shortstack{Cell structure\\ and function} & & \shortstack{British \\romanticism} & \shortstack{Literary and \\cultural theory} & \shortstack{Stylistics \\~\textbf{~\textcolor{red}{in literature}}} \\
    \cline{2-4}
    \cline{6-8}
    & \multicolumn{3}{c}{\shortstack{Principles of genetics}} & & \multicolumn{3}{c}{\shortstack{Aspects of modernism ~\textbf{~\textcolor{red}{in literature}}}}  \\
    \hline
    \hline
    \multirow{2}{*}[-4em]{\textbf{Chemistry}} & \shortstack{Solid state \\ chemistry} & \shortstack{Shapes, properties and \\ reactions of molecules} & \shortstack{Organic and \\ biological chemistry} & \multirow{2}{*}[-4em]{\textbf{Politics}} & \shortstack{Central themes \\in political thought} & \shortstack{Modern British  \\ politics} & \shortstack{Capital labour and power: \\ Britain 1707-1939}\\
    \cline{2-4}
    \cline{6-8}
    & \shortstack{Chemistry for the \\ physical sciences} & \shortstack{Molecular \\ pharmacology} & \shortstack{States of \\ matter~\textbf{~\textcolor{red}{in chemistry}}} & & \shortstack{The holocaust \\ ~\textbf{~\textcolor{red}{in politics}}} & ~\textbf{~\textcolor{red}{\shortstack{War in the industrial \\ age politics~\tablefootnote{Total War in the modern era}}}} & \shortstack{Freedom, power and resistance: \\ an introduction to political ideas} \\
    \cline{2-4}
    \cline{6-8}
    & \shortstack{Chemistry of \\ materials} & \shortstack{Inorganic \\ chemistry} & \shortstack{The global \\ Earth system} & & \shortstack{International \\politics} & \shortstack{Making of the modern \\ world~\textbf{~\textcolor{red}{in politics}}} & \shortstack{The political \\ economy of development} \\
    \cline{2-4}
    \cline{6-8}
    & \multicolumn{3}{c}{\shortstack{Mineralogy and petr\textbf{\textcolor{red}{o}}logy} (typo exists in the original query)}  & & \multicolumn{3}{c}{\shortstack{Comparing extremism in European liberal democracies}} \\
    \hline
    \hline
    \multirow{2}{*}[-4em]{\textbf{\shortstack{Computer \\ Science}}} & \shortstack{Organisational behaviour \\ in practice} & \shortstack{Principles of \\ programming} & \shortstack{Data management in \\ computer  science} & \multirow{2}{*}[-4em]{\textbf{Psychology}} & \shortstack{Cell biology \\in psychology} & \shortstack{Mind and behaviour} & \shortstack{Exploring effective \\ learning ~\textbf{~\textcolor{red}{in psychology}}}\\
    \cline{2-4}
    \cline{6-8}
    & \shortstack{Mathematics for \\ computer science} & \shortstack{Languages and \\ computability} & \shortstack{Fundamentals of \\~\textbf{~\textcolor{red}{Computer}} Design~\tablefootnote{Fundamentals of Design}} & & \shortstack{Experimental methods and \\ statistical} &  \shortstack{Individual and \\ social processes} & \shortstack{Development \\ psychology} \\
    \cline{2-4}
    \cline{6-8}
    & \shortstack{~\textbf{~\textcolor{red}{Personal Computer}} \\ technology~\tablefootnote{PC technology}} & \shortstack{Image \\ processing} & \shortstack{Software systems development \\~\textbf{~\textcolor{red}{in computer science}}} & & \shortstack{Brain and \\cognition~\textbf{~\textcolor{red}{in psychology}}} & \shortstack{Social psychology} & \shortstack{Humans in biological \\ perspective~\textbf{~\textcolor{red}{in psychology}}} \\
    \cline{2-4}
    \cline{6-8}
    & \multicolumn{3}{c}{\shortstack{Human computer interaction}} & & \multicolumn{3}{c}{\shortstack{Evolution and behaviour~\textbf{~\textcolor{red}{in psychology}}}} \\
    \hline
    \hline
    \multirow{2}{*}[-4em]{\textbf{\shortstack{Mathematics}}} & \shortstack{Calculus} & \shortstack{Algebra} & \shortstack{Structured \\ programming} & \multirow{2}{*}[-4em]{\textbf{\shortstack{Public \\ Relations}}} & \shortstack{Business \\ strategy} & \shortstack{Internal \\ corporate communication} & \shortstack{Social media \\ or public relations}\\
    \cline{2-4}
    \cline{6-8}
    & \shortstack{Algorithms \\ and applications} & \shortstack{Coordinate and \\ vector geometry} & \shortstack{Differential \\ equations} & & \shortstack{Work and \\ organisational change} &  ~\textbf{~\textcolor{red}{\shortstack{Behavioural \\ science~\tablefootnote{Human behaviour}}}} & \shortstack{Management \\ in context} \\
    \cline{2-4}
    \cline{6-8}
    & \shortstack{Probability} & \shortstack{Regression \\ and anova} & \shortstack{Analytical and computational \\ foundations~\textbf{~\textcolor{red}{in maths}}} & & ~\textbf{~\textcolor{red}{\shortstack{Work experience \\ in public relations~\tablefootnote{Transition to work}}}} & \shortstack{Business \\ fundamentals} & \shortstack{Managing \\ the brand} \\
    \cline{2-4}
    \cline{6-8}
    & \multicolumn{3}{c}{\shortstack{Problem solving methods~\textbf{~\textcolor{red}{in maths}}}} & & \multicolumn{3}{c}{\shortstack{Design in marketing}} \\
    \hline
    \hline
    \multirow{2}{*}[-4em]{\textbf{\shortstack{Physics}}} & \shortstack{Laboratory \\ physics} & \shortstack{Contemporary \\ physics} & \shortstack{Mathematical techniques \\ ~\textbf{~\textcolor{red}{in physics}}} & \multirow{2}{*}[-4em]{\textbf{\shortstack{Sociology}}} & \shortstack{Observing \\ in sociology} & \shortstack{Urban \\ sociology} & \shortstack{Understanding deviance and social \\ problems~\textbf{~\textcolor{red}{in sociology}}}\\
    \cline{2-4}
    \cline{6-8}
    & \shortstack{Quantum \\ physics} & \shortstack{Newtonian and \\ relativistic mechanics} & \shortstack{Fabric of \\ physics} & & \shortstack{Individual and \\ society} & \shortstack{Applied \\ ethics} & \shortstack{Media and \\ crime~\textbf{~\textcolor{red}{in sociology}}} \\
    \cline{2-4}
    \cline{6-8}
    & \shortstack{Plasma and \\ fluids~\textbf{~\textcolor{red}{in physics}}} & \shortstack{Special and \\ general relativity} & \shortstack{Analysing the \\ nanoscale and magnetism} & & \shortstack{Nature and \\ society~\textbf{~\textcolor{red}{in sociology}}} & \shortstack{Sexuality and \\ social control~\textbf{~\textcolor{red}{in sociology}}} & \shortstack{Contemporary work and \\ organisational life~\textbf{~\textcolor{red}{in sociology}}} \\
    \cline{2-4}
    \cline{6-8}
    & \multicolumn{3}{c}{\shortstack{Stellar physics}} & & \multicolumn{3}{c}{\shortstack{Mobilisation, social movements and protest~\textbf{~\textcolor{red}{in sociology}}}}\tabularnewline
    \hline
\label{tab:dataset_new}
\end{tabular}}
%\vspace{-2.3em}
\end{table}

\section{Experimental Setup}
\label{sec:experimental}
In this section we provide a description of our experimental setup based on the proposed methodology as defined in Section~\ref{sec:gender_framework}. 
We initially provide information about the dataset and the annotation process. Then, we show our~\emph{perceived} gender bias results and discuss them.

\subsection{Dataset Preparation}
\label{sec:dataset}
In this work, we aim to mimic a user scenario in which the user is trying to decide on his/her major through searching educational queries or course modules of various majors on YouTube. Thus, we designed our study accordingly and obtained all the educational queries from~\emph{TheUniGuide}. The first reason why we chose TheUniGuide is that when we searched a query of "university chemistry courses" in~\emph{incognito mode} with a UK proxy to construct the set of educational queries for~\emph{chemistry}, TheUniGuide appears as the top result in Google search. Second, the other search results were mainly the official pages of different universities about the corresponding major and we did not want to select the curriculum information of a specific university. Third, we examined the webpages for different majors from STEM and NON-STEM fields and observed that TheUniGuide website provides comprehensive information about a major~\footnote{For the major of "chemistry", please go to~\url{https://www.theuniguide.co.uk/subjects/chemistry}.}.

We selected 5 STEM majors which are chemistry, physics, biology, maths, and computer science,  and 5 NON-STEM majors which are sociology, psychology, politics, public relations, and English language and literature. We selected these majors since we believe that they  span  distinct areas in STEM and NON-STEM fields which might have different male/female gender proportions. 
%For the crawling, we simulated a specific scenario in which the user is a prospective university student who uses YouTube for deciding on his/her major in STEM and NON-STEM fields. 
Initially, we crawled the course modules of each selected major from TheUniGuide. We used a UK proxy, the YouTube desktop version in~\emph{incognito mode} and the region was set as UK, language as English automatically and the other settings were left as default. 
%Note that in the filter options at the top, "sort by" option was by default selected as "relevance" which means that the search results will be ranked based on relevance.
%
In the scope of this work, since personalised search might complicate the bias analysis, we decided to design our analysis in unpersonalised search settings. In our initial design, we included top-12 relevant videos and recommended videos by YouTube. However, we observed that the recommended video results were not context-specific about the issued query. Thus, we changed the data crawling process, decided to crawl only relevant~\emph{YVRP}s for a given query and obtained all the course modules from TheUniGuide to use them as user queries.
Note that TheUniGuide has 10 course modules for each major, thus each STEM and NON-STEM fields has 50 queries and 50 ranked lists in total. Since we only crawled the top 12 relevant~\emph{YVRP}s per query, for each course module/query we obtained 12 video URLs and for each major we obtained 120, thus crawling the 1200 video URLs in total.

\begin{table}[!t]
    \centering
    \caption{
    \emph{Perceived} gender bias in YouTube for the $top-12$ relevant results, p-values of a two-tailed paired t-test computed between STEM and NON-STEM fields
    %P-values of two-tailed paired t-test computer between engine 1 and 2.
    %$\ast$ indicates non statistical significance %while $\dag$ (p-value < 0.001) statistical significance 
    %of a one-sample t-test.
    }
    \setlength{\tabcolsep}{2.8pt}
    \renewcommand{\arraystretch}{1.1}
    \begin{tabular}{cccccccc|ccccccccc}
        \hline\hline
        & & $\mathcal{R}ep@3$  & & $\mathcal{R}ep@6$  & & $\mathcal{R}ep@12$  && $\mathcal{E}xp@3$ && $\mathcal{E}xp@6$  && $\mathcal{E}xp@12$  & \\
        \hline
        \multirow{3}{*}{MB} 
        & STEM & 0.6200*** && 0.5460*** && 0.5558*** && 0.6012*** && 0.5541*** && 0.5600*** &\\	
        & NON-STEM & 0.3067*** && 0.2820*** && 0.3266*** && 0.3083*** && 0.2880*** && 0.3144*** &\\
        \cline{2-14}
        & p-value & $ .004$ & & $ .005$ & & $ .005$ && $.011$ && $.005$ && $.002$ &\\	
        & effect size d & $ 0.59$ & & $ 0.58$ & & $ 0.58$ && $0.53$ && $0.58$ && $0.63$ &\\	
        \hline
        \multirow{3}{*}{MAB} 
        & STEM & 0.7000***	&& 0.5940*** &&	0.5794*** && \textbf{0.7136***} && 0.6225*** && \textbf{0.5877}*** &\\	
        & NON-STEM & 0.5467*** &&0.4767*** &&	0.4808*** && 0.5566*** && 0.4683*** && 0.4449***&\\
        \cline{2-14}
        & p-value & $ .035$ & & $ .077$ & & $ .086$ && $.033$ && $.017$ && $.017$ &\\	
        & effect size d & $ 0.43$ & & $ 0.36$ & & $ 0.35$ &&  $ 0.44$ && $ 0.49$&& $ 0.49$ &\\	
        \hline\hline
    \end{tabular}
    \label{tab:gendernew_real}
\end{table}

Additionally, for some queries YouTube did not return relevant video results in the educational context, therefore we slightly modified the queries/course modules as highlighted by red color in Table~\ref{tab:dataset_new}. For the majority of the queries, we took them as they are written in TheUniGuide with lowercase/uppercase letters as well as punctuation symbols to avoid injecting our personal bias. For the rest, if we observed that YouTube did not return context-specific video results in response to the original query, we slightly modified the original query solely by adding the context/major field information as displayed in Table~\ref{tab:dataset_new}. If this solution was not sufficient, then we had to change the query itself to specify the context properly. For instance, for the original query of~\emph{Human behaviour}, YouTube returned the music videos of Björk, an Icelandic singer, therefore we paraphrased those queries. We first collected sufficient information on the Web to change those queries properly to obtain more context-specific results. In this way, we believe that our search scenario became more realistic in the educational context which can help us to better detect~\emph{perceived} gender bias that the user is exposed to in real world. For the sake of reproducibility, the annotated dataset is publicly available at~\url{https://github.com/gizem-gg/Youtube-Gender-Bias}.

\begin{table*}[!t]
    \centering
    \caption{
    \emph{Perceived} gender bias in YouTube for the $top-12$ relevant results, p-values of a two-tailed paired t-test computed between STEM and NON-STEM fields
    %P-values of two-tailed paired t-test computer between engine 1 and 2.
    %$\ast$ indicates non statistical significance %while $\dag$ (p-value < 0.001) statistical significance 
    %of a one-sample t-test.
    }
    \setlength{\tabcolsep}{2.8pt}
    \renewcommand{\arraystretch}{1.1}
    \begin{tabular}{cccccc|cccc|ccccc}
        \hline\hline
        & & STEM  & & NON-STEM  & & & STEM & NON-STEM & & & STEM &  NON-STEM & \\
        \hline
        \multirow{6}{*}{\textbf{MB}}
        & $\mathcal{R}ep@3$ & 0.6200*** && 0.3067*** && $\mathcal{R}ep@6$ & 0.5460*** & 0.2820*** && 
        $\mathcal{R}ep@3$ & 0.6200*** & 0.3067*** &\\	
        
        & $\mathcal{R}ep@6$ & 0.5460*** && 0.2820*** && $\mathcal{R}ep@12$ & 0.5558*** & 0.3266*** && 
        $\mathcal{R}ep@12$ & 0.5558*** &0.3266*** &\\
        \cline{2-14}
        & p-value & $ .112$ & & $ .619$ & & & $ .781$ & $ .199 $ &&& 
        $ .259 $ & $ .765$ &\\
        \cline{2-14}
        %\\
        %\multirow{3}{*}{MAB} 
        & $\mathcal{E}xp@3$ & 0.6012*** && 0.3083*** && $\mathcal{E}xp@6$ & 0.5541*** & 0.2880*** && 
        $\mathcal{E}xp@3$ & 0.6012*** & 0.3083*** &\\	
        
        & $\mathcal{E}xp@6$ & 0.5541*** && 0.2880*** && $\mathcal{E}xp@12$ & 0.5600*** & 0.3144*** && 
        $\mathcal{E}xp@12$ & 0.5600*** &0.3144*** &\\
        \cline{2-14}
        & p-value & $ .191 $ & & $ .591 $ & & & $ .829 $ & $ .313 $ &&&
        $ .391 $ & $ .911 $ &\\		
        \hline\hline
    \end{tabular}
    \label{tab:gendernewcutoff_MB}
\end{table*}

\begin{table*}[!t]
    \centering
    \caption{
    \emph{Perceived} gender bias in YouTube for the $top-12$ relevant results, p-values of a two-tailed paired t-test computed between STEM and NON-STEM fields
    %P-values of two-tailed paired t-test computer between engine 1 and 2.
    %$\ast$ indicates non statistical significance %while $\dag$ (p-value < 0.001) statistical significance 
    %of a one-sample t-test.
    }
    \setlength{\tabcolsep}{3.3pt}
    \renewcommand{\arraystretch}{1.3}
    \begin{tabular}{cccccc|cccc|ccccc}
        \hline\hline
        & & STEM  & & NON-STEM  & & & STEM & NON-STEM & & & STEM &  NON-STEM & \\
        \hline
        \multirow{6}{*}{\textbf{MAB}}
        & $\mathcal{R}ep@3$ & 0.7000*** && 0.5467*** && $\mathcal{R}ep@6$ & 0.5940*** & 0.4767*** && 
        $\mathcal{R}ep@3$ & 0.7000*** & 0.5467*** &\\	
        
        & $\mathcal{R}ep@6$ & 0.5940*** && 0.4767*** && $\mathcal{R}ep@12$ & 0.5794*** & 0.4808*** && 
        $\mathcal{R}ep@12$ & 0.5794*** &0.4808*** &\\
        \cline{2-14}
        & p-value & $ .016$ & & $ .112 $ & & & $ .663 $ & $ .887 $ &&& 
        $ .013 $ & $ .216 $ &\\
        \cline{2-14}
        %\\
        %\multirow{3}{*}{MAB} 
        & $\mathcal{E}xp@3$ & 0.7136*** && 0.5566*** && $\mathcal{E}xp@6$ & 0.6225*** & 0.4683*** && 
        $\mathcal{E}xp@3$ & \textbf{0.7136***} & 0.5566*** &\\	
        
        & $\mathcal{E}xp@6$ & 0.6225*** && 0.4683*** && $\mathcal{E}xp@12$ & 0.5877*** & 0.4449*** && 
        $\mathcal{E}xp@12$ & \textbf{0.5877***} &0.4449*** &\\
        \cline{2-14}
        & p-value & $ .007 $ & & $ .013 $ & & & $ .150 $ & $ .362 $ &&& 
        $ \textbf{.001} $ & $ .021 $ &\\		
        \hline\hline
    \end{tabular}
    \label{tab:gendernewcutoff_MAB}
\end{table*}

\begin{table*}[!t]
    \centering
    \caption{
    \emph{Perceived} Gender bias for specific majors of STEM and NON-STEM fields in YouTube for the $top-12$ relevant results
    %P-values of two-tailed paired t-test computer between engine 1 and 2.
    %$\ast$ indicates non statistical significance %while $\dag$ (p-value < 0.001) statistical significance 
    %of a one-sample t-test.
    }
    \setlength{\tabcolsep}{2.4pt}
    \renewcommand{\arraystretch}{1.3}
    \begin{tabular}{cccccc||cccccc}
        \hline\hline
        & $Biology$  & $Chemistry$ &$CS$ & $Maths$ &$Physics$ & $Eng.\:Lan.\:Lit.$  &$Politics$ &$Psychology$ &$Pub.\:Rel.$ &$Sociology$ & \\
        \hline
        $\mathcal{R}ep@3$
        & \textbf{\textcolor{blue}{0.4333}} & 0.5667 & \textbf{\textcolor{red}{0.7667}} & 0.7333 & 0.5667 & \textbf{\textcolor{red}{0.6333}} & 0.3667 & 0.3000 & \textbf{\textcolor{blue}{-0.1667}} & 0.4333 \\
        $\mathcal{R}ep@6$
        & \textbf{\textcolor{blue}{0.2833}} & 0.5033 & 0.5533 & \textbf{\textcolor{red}{0.7667}} & 0.6067 & \textbf{\textcolor{red}{0.7033}} & 0.3833 & 0.3400 & \textbf{\textcolor{blue}{-0.1667}} & 0.1667 &\\
        $\mathcal{R}ep@12$
        & \textbf{\textcolor{blue}{0.2061}} & 0.5699 & 0.5930 & 0.6833 & \textbf{\textcolor{red}{0.7098}}& \textbf{\textcolor{red}{0.7722}} & 0.3701 & 0.3619 & \textbf{\textcolor{blue}{-0.0466}} & 0.1809 &\\
        \Xhline{2.5\arrayrulewidth}
        $\mathcal{E}xp@3$
        & \textbf{\textcolor{blue}{0.4524}} & 0.5235 & 0.7366 & \textbf{\textcolor{red}{0.7631}} & 0.5007 & \textbf{\textcolor{red}{0.5831}} & 0.3765 & 0.2656 & \textbf{\textcolor{blue}{-0.1696}} & 0.5088 & \\
        $\mathcal{E}xp@6$
        & \textbf{\textcolor{blue}{0.3523}} & 0.4846 & 0.5858 & \textbf{\textcolor{red}{0.7762}} & 0.5527 & \textbf{\textcolor{red}{0.6446}} & 0.3857 & 0.3022 & \textbf{\textcolor{blue}{-0.1681}} & 0.2908 & \\
        $\mathcal{E}xp@12$
        & \textbf{\textcolor{blue}{0.2736}} & 0.5451 & 0.6031 & \textbf{\textcolor{red}{0.7119}} & 0.6478 & \textbf{\textcolor{red}{0.7138}} & 0.3781 & 0.3241 & \textbf{\textcolor{blue}{-0.0917}} & 0.2560 &\\
        \hline
        \hline
    \end{tabular}
    \label{tab:gender_majorarea}
\end{table*}

As for the annotation procedure, a video in a~\emph{YVRP} is annotated based on its relevancy with respect to the given query as \emph{relevant}, \emph{not-relevant}, or \emph{N/A}. If the video is \emph{relevant} to the given query, then its narrators' gender perceived by the viewer is annotated using the gender labels of \emph{male}, \emph{female}, or \emph{neutral}. Before the annotation of the actual dataset, as two annotators we initially annotated the dataset of top-12 relevant~\emph{YVRP}s in which we crawled for the first user study. Then, we modified the design of our study and, thus the data crawling procedure. Nonetheless, we computed inter-rater agreement score which is calculated as a percentage of agreement between two annotators. We examined pairwise agreement; enter 1 if there is agreement and 0 if there is no agreement. After that, we calculated the mean of the fractions and the inter-rater agreement score for our initial study on the top-12 relevant~\emph{YVRP}s was over 0.90. Since the inter-rater agreement score is high, it shows that the annotation procedure does not prone to disagreements due to the simplicity of the task and does not require expert knowledge. Thus, the labelling has been fulfilled using a single annotator. Conditions of annotating a video with these labels are detailed in Section~\ref{sec:prelim}.

\begin{figure}[!t]
    \centering
    \setcounter{figure}{1}
    \includegraphics[width=0.6\textwidth]{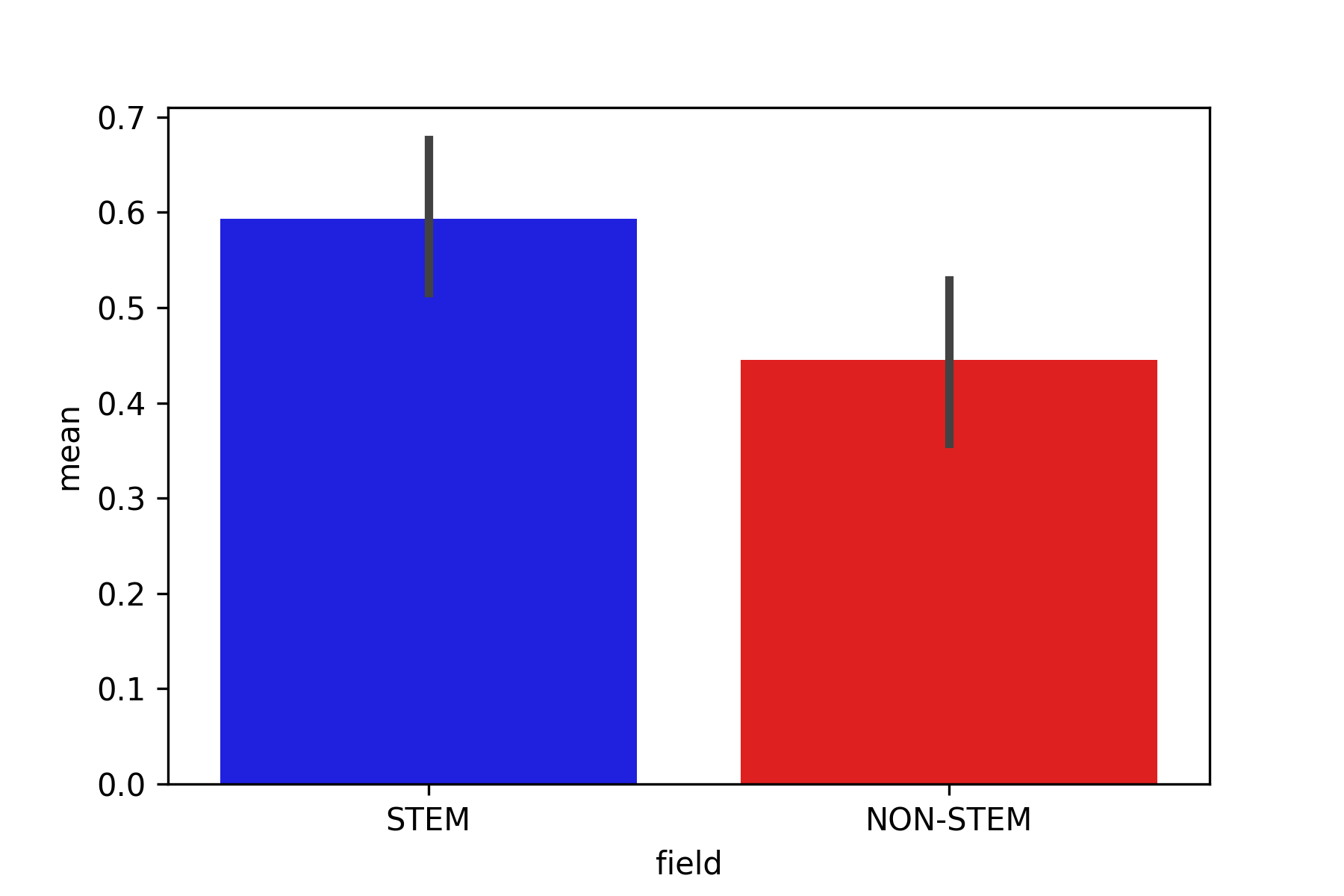}
    \caption{MB scores of $\Delta_{\mathcal{E}xp@12}$ measured on~\emph{perceived} gender labels of STEM and NON-STEM fields.}
    \label{fig:exp12_barplot}
\end{figure}

\subsection{Results \& Discussion}
\label{sec:results}
Initially, we verify if the~\emph{YVRP}s are biased using our representation and exposure measures (\textbf{RQ1}). If so, we then investigate if the~\emph{YVRP}s returned in response to the educational queries of STEM and NON-STEM fields suffer from a different level of bias (\textbf{RQ2}) by examining if the difference between the bias scores of the corresponding~\emph{YVRP}s are statistically significant. In Table~\ref{tab:gendernew_real}, all MB scores are positive and regarding the \textbf{RQ1}, the~\emph{YVRP}s of STEM and NON-STEM fields are both biased. We applied the one-sample t-test on MB scores to check the existence of bias and these biases are statistically significant -- there exists a systematic gender bias, i.e. preference of one gender over another, with p-value < .001. These results indicate that both STEM and NON-STEM fields are biased towards the male gender (all MB scores are positive). Based on MAB scores, we can observe that the~\emph{YVRP}s of both STEM and NON-STEM fields suffer from an absolute bias. Regarding the~\textbf{RQ3}, we first investigate if different cut-off values have an impact on the existence of bias and further if those cut-off values affect the magnitude of bias difference between STEM and NON-STEM fields. These findings indicate that both STEM and NON-STEM fields are biased regardless of the cut-off values, yet cut-off values might impact on the magnitude of bias difference between the two fields. Lastly, we further investigate the effect of different cut-off values on the magnitude of bias in the same field of STEM or NON-STEM~\textbf{(RQ4)}.

In Table~\ref{tab:gendernew_real}, the difference between the MB scores of two fields is shown to be statistically significant with the two-tailed independent t-test for the measures $\mathcal{R}ep@n$ and $\mathcal{E}xp@n$ regardless of the cut-off value and in terms of the magnitude, $\mathcal{E}xp@12$ gives the highest difference in bias with the effect size of $0.63$ (\textbf{RQ2}). Unlike MBs, the difference between the MAB scores of two fields is shown to be statistically significant on $\mathcal{R}ep@3$, i.e. top 3 videos returned, with  p-value = $0.035$ and statistically not significant on $\mathcal{R}ep@6$ and $\mathcal{R}ep@12$. On the other hand, the difference between the MAB scores of two fields is shown to be statistically significant on the exposure measure across all three cut-off values with p-value = $0.033$, $0.017$, $0.017$ respectively for $n$ = $3$, $6$, $12$ (\textbf{RQ2}).
These findings about MABs and the different user search scenarios that are modelled by the two bias evaluation measures suggest that the~\emph{perceived} gender bias to which the users are exposed may change based on their behaviour. A user that always reviews the first 6 or 12 results (as modelled by $\mathcal{R}ep@6$ and $\mathcal{R}ep@12$) might perceive the same gender bias between the~\emph{YVRP}s of STEM and NON-STEM fields. Yet, a less systematic user, which just reviews the top results, might perceive that STEM field is more biased than NON-STEM field -- see the MAB scores of $\mathcal{R}ep@3$ and $\mathcal{E}xp@n$ for $n$ = $3$, $6$, $12$ in Table~\ref{tab:gendernew_real} (\textbf{RQ3}). Note that the exposure measure generally shows a higher difference in bias with bigger effect sizes since it is more sensitive to rank information. In addition to these, unlike MAB scores we observe significant difference in MB scores of STEM and NON-STEM fields across all measures with higher difference in magnitude (bigger effect sizes). These empirical findings imply that the NON-STEM field is biased for some educational queries towards the male gender and for others towards the female gender -- they possibly cancelled each other out. Although the same situation applies to the STEM field, it seems that STEM has less number of queries that are biased towards the female gender -- MB and MAB scores of STEM field are more similar.

In Table~\ref{tab:gendernewcutoff_MB} and~\ref{tab:gendernewcutoff_MAB}, we investigate if different cut-off values affect the magnitude of~\emph{perceived} bias in STEM and NON-STEM fields for the MBs and MABs of representation and exposure measures~\textbf{(RQ4)}. In Table~\ref{tab:gendernewcutoff_MB}, we observe that magnitude of bias does not differ with various cut-off values on the MB scores. The differences between the MB scores of each field with different cut-off values are shown to be statistically not significant with the two tailed paired t-test. Similarly, in Table~\ref{tab:gendernewcutoff_MAB}, the differences between the MAB scores are statistically not significant, except the difference between the MABs of STEM field for the measures $\mathcal{E}xp@3$ and $\mathcal{E}xp@12$ which is statistically significant, i.e. p-value < $0.05$ due to Bonferroni correction.

\begin{figure}[!t]
\centering
\caption{MAB scores of $\Delta_{\mathcal{E}xp@n}$ measured on~\emph{perceived} gender labels of STEM field.}
\captionsetup{justification=centering}
  \begin{subfigure}[b]{0.45\textwidth}
    \includegraphics[width=\linewidth]{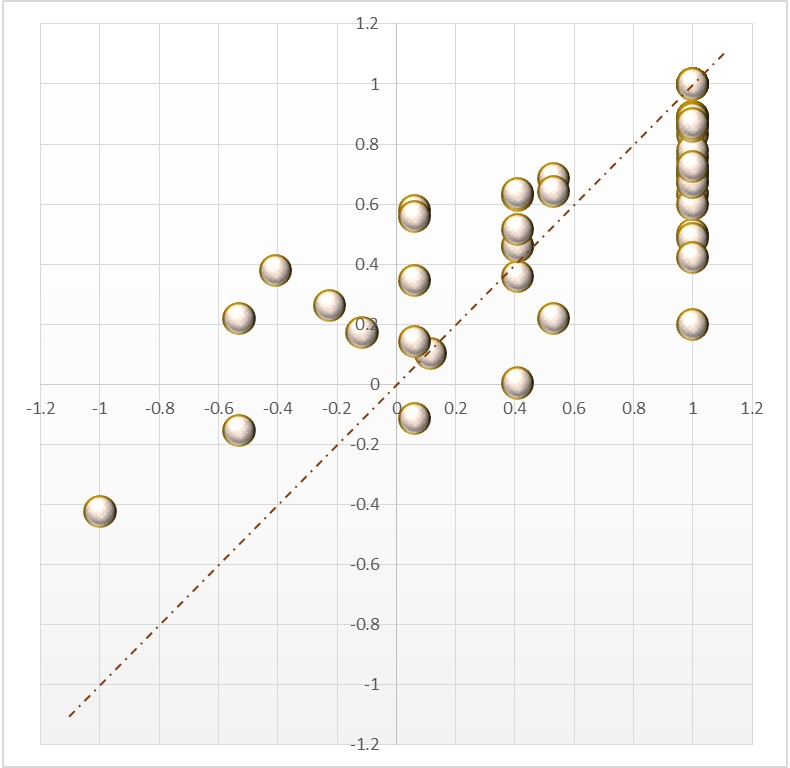}
   \caption{$\Delta_{\mathcal{E}xp@n}$ measured on~\emph{perceived} gender labels, where x-axis is $n = 3$ and y-axis is $n = 12$.}
    \label{fig:deltabias_3_12}
  \end{subfigure}
  %
  %\hfill
  \hspace{0.04\textwidth}
  \begin{subfigure}[b]{0.45\textwidth}
    \includegraphics[width=\linewidth]{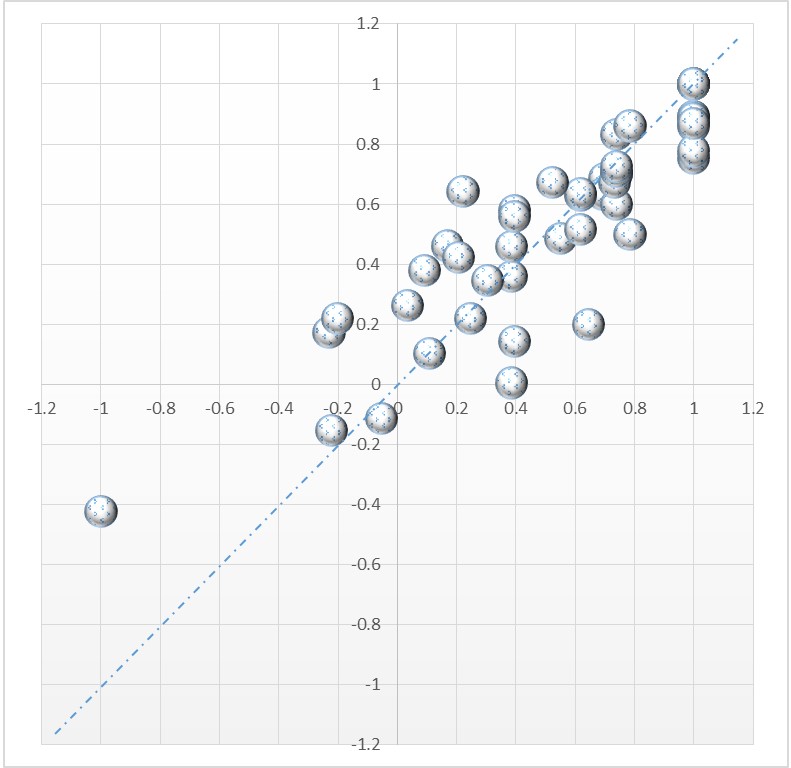}
    \caption{$\Delta_{\mathcal{E}xp@n}$ measured on~\emph{perceived} gender labels, where x-axis is $n = 6$ and y-axis is $n = 12$.}
    \label{fig:deltabias_6_12}
  \end{subfigure}
  \label{fig:delta_bias_stem}
\end{figure}

Lastly, in Table~\ref{tab:gender_majorarea} we investigate which majors show the highest/lowest bias in STEM and NON-STEM fields. The empirical findings indicate that~\emph{Biology} is biased on both measures towards the male gender which shows the lowest bias score in STEM field and different cut-off values do not affect this. Unlike~\emph{Biology}, different measures and cut-off values change the major which shows the highest bias score towards the male gender. For the exposure measure,~\emph{Mathematics (a.k.a Maths)} is also biased towards the male gender and shows the highest bias score regardless of the cut-off values. On the other hand, for the representation measure the major with the highest bias score depends on the cut-off value. For $n$ = $3$,~\emph{Computer Science (a.k.a CS)} is the major showing the highest bias score, while for $n$ = $6$,~\emph{Mathematics (a.k.a Maths)} as in the case of exposure measures, and for the full list~\emph{Physics} shows the highest score, where $n$ = 12. Nonetheless for the NON-STEM field, the majors showing highest/lowest bias scores change neither with different measures nor cut-off values. Among the majors in NON-STEM field,~\emph{Public Relations (a.k.a Pub. Rel.)} shows the lowest bias which is the only major that is biased towards the female gender, whereas ~\emph{English Language and Literature (a.k.a Eng. Lan. Lit.)} shows the highest bias score which is biased towards the male gender like majority of the majors in STEM and NON-STEM fields. 

Moreover, the majors showing the highest bias in STEM field seem to be more biased in absolute value (magnitude) on average than their counterparts in NON-STEM field. Similarly,~\emph{Biology} shows higher bias scores on average than~\emph{Public Relations (a.k.a Pub. Rel.)} when we compare the absolute value of their scores. These results seem to be consistent with our aforementioned findings in Table~\ref{tab:gendernew_real} that STEM field is more biased which is towards the male gender. Also, these empirical findings in Table~\ref{tab:gender_majorarea} support our implication that the~\emph{YVRP}s of some majors in NON-STEM field are biased towards the male, while others towards the female gender. Furthermore, the bias scores of majors in STEM field are more similar -- standard deviation is smaller than the majors in NON-STEM field, i.e. average standard deviation is $0.15$ for STEM majors while $0.27$ for NON-STEM. Finally, looking at the major-specific bias scores we can observe that there exists a noticeable bias towards the male gender, even if~\emph{Public Relations (a.k.a Pub. Rel.)} seems to be biased towards the female gender, yet unlike STEM majors it does not show a strong bias.

The difference in magnitude of bias between STEM and NON-STEM fields is illustrated in Figure~\ref{fig:exp12_barplot} with the visual comparison of MBs for the measure $\Delta_{\mathcal{E}xp@12}$ since it shows the maximal difference with an effect size of $0.63$ (please refer to Table~\ref{tab:gendernew_real}). Figure~\ref{fig:exp12_barplot} also supports the conclusion that the overall results are biased towards the male gender and the STEM field is more biased. Moreover, the results are displayed in Figure~\ref{fig:delta_bias_stem}~(a) and (b) which refer to the values of STEM field using MABs of exposure ($\Delta_{\mathcal{E}xp@n}$), where $n = 3$ vs $n = 12 $ and $n = 6$ vs $n = 12$ respectively. We deliberately display the MAB scores of $\Delta_{\mathcal{E}xp@n}$ in STEM field for $n = 3$ and $n = 12$ cut-off values since this is the only difference that is statistically significant in Table~\ref{tab:gendernewcutoff_MAB}. In those figures, positive coordinates denote the $\Delta_{\mathcal{E}xp@n}$ scores of the~\emph{YVRP}s of queries annotated more with male gender and negative coordinates more with the female gender for two different cut-off values in the STEM field. 
Similarly, Figure~\ref{fig:delta_bias_stem}~(a) and (b) confirm our findings that the overall results are biased towards the male gender (for the STEM) -- majority of the queries are in the upper right quadrant in both figures. There are trendlines ($y = x$) in both of these figures and Figure~\ref{fig:delta_bias_stem}~(b) displays more similar~\emph{perceived} bias scores in terms of magnitude for $n = 6$ vs $n = 12$, i.e. bubbles are mostly around the trendline. Yet, in Figure~\ref{fig:delta_bias_stem}~(a) we observe more dispersed bubbles; they are not gathered around the trendline which means that the magnitude of bias differs between $n = 3$ and $n = 12$. Both Figure~\ref{fig:delta_bias_stem}~(a) and (b) confirm our findings in Table~\ref{tab:gendernewcutoff_MAB}.

\section{Conclusion \& Future Work}
\label{sec:conclusion}
In this work, we investigated possible bias in educational videos with a case study on YouTube. In order to achieve that, we propose two new measures of bias that are suitable for the group-fairness criteria of~\emph{equality of outcome}. We applied these measures first to investigate the existence of~\emph{perceived} gender bias in the~\emph{YVRP}s in response to STEM and NON-STEM educational queries. Then, we examined if there is a significant difference in bias between STEM and NON-STEM fields. Lastly, we inspected the impact of different cut-off values on the bias results, i.e. on the existence of bias, difference in magnitude between the two fields, as well as in the same field. Our initial results show that both STEM and NON-STEM fields are biased towards the male gender. Also, the empirical findings indicate that STEM and NON-STEM fields generally show significantly different magnitude of~\emph{perceived} gender bias from each other towards the educational queries. Nonetheless, we observed that different bias measures as well as cut-off values might affect the magnitude of bias difference between STEM and NON-STEM fields. Nonetheless, we observed that different cut-off values might also impact on the magnitude of bias that a field shows. In this work, we intended to analyse~\emph{YVRP}s without the effect of personalisation. Therefore, these results point out that there exists~\emph{perceived} gender bias in~\emph{YVRP}s that could be observed even with a crude analysis and even though the personalisation effect is minimized. These findings suggest that the~\emph{YVRP}s are biased as displayed irrespective of the user's YouTube search history, i.e. for instance there is no space for the following validation, clicking the educational videos mostly with male narrators lead to these results. We believe that this preliminary study is mainly valuable due to the following three reasons. First, this study views YouTube as an online educational platform which is the reality in practice in addition to the platforms that only serve as online educational websites. Second, this work investigates gender bias based on the~\emph{perceived} gender of the narrator by evaluating the gender bias in the context of search, by taking inspiration from widely-used IR measures. Third, gender bias detection in online educational systems is critical, especially by simulating a real user scenario to crawl for the bias analysis since the consequences may be severe; these platforms exacerbate existing biases connected to gender inequality in society.

We note that although the main focus of this study is not to investigate the source of bias, i.e. algorithmic or corpus bias, we believe that our results can be seen as a potential indicator. One possible future direction is to investigate a non-binary~\emph{perceived} gender bias analysis. Another one is to focus on the source of bias via an automatic model to detect narrators'~\emph{perceived} gender. Alternative future direction could be to include personalisation in our analysis, i.e. by analysing gender bias from female and male users' perspective to investigate if gender of the user affects the existence and/or magnitude of bias. Lastly, we can work on simple and efficient algorithms that could run in large-scale systems so even if the bias comes from the corpus, those platforms can somewhat mitigate the bias.

\newpage
%\section*{Acknowledgements}
%Funding: This research did not receive any specific grant from funding agencies in the public, commercial, or not-for-profit sectors.

\printcredits

%\bibliographystyle{cas-model2-names.bst}\biboptions{authoryear}
\bibliography{main}
\bibliographystyle{cas-model2-names}
%\biboptions{authoryear}


\begin{comment}
\bio{}
\endbio

\bio{pic1}
\endbio

\end{comment}

\end{document}